\documentclass[vecphys]{svmult}

\usepackage{makeidx}         % allows index generation
\usepackage{graphicx}        % standard LaTeX graphics tool
\usepackage{multicol}        % used for the two-column index
\usepackage{cite}            % adjusts the "syntax" of the refs in the
\usepackage[bottom]{footmisc}% places footnotes at page bottom
\usepackage{amsmath,amssymb}

\makeindex             % used for the subject index
\begin{document}

\title{Resonant-State Expansion of the Fano Peak in Open Quantum Systems}
\titlerunning{Resonant-State Expansion of the Fano Peak}
\author{Naomichi Hatano \and Gonzalo Ordonez}
%\authorrunning{Naomichi Hatano \and Gonzalo Ordonez}
\institute{Institute of Industrial Science, The University of Tokyo, 5-1-5 Kashiwanoha, Kashiwa, Chiba 277-8574, Japan
\texttt{hatano@iis.u-tokyo.ac.jp}
\and Department of Physics and Astronomy, Butler University, 4600 Sunset Avenue, Indianapolis, Indiana 46208, U.S.A. \texttt{gordonez@butler.edu}}
% Use the package "url.sty" to avoid problems with special characters
% used in your e-mail or web address.
% Addresses should be removed from contribution and entered into
% blist.tex" (by the compiler).

\maketitle

\begin{abstract}
We describe the Fano asymmetry by expanding the transmission amplitude with respect to states with point spectra (discrete eigenstates), including not only bound states but also resonant states with complex eigenvalues. We first introduce a novel complete set that spans the Hilbert space of the central part of an open quantum-dot system. This complete set contains all states of point spectra, but does not contain any states of continuous spectra. We thereby analytically expand the conductance of the dot in terms of all discrete states without any background integrals. 
This expansion implies that the resonant states produce the main contributions to the electron transmission.

We then explain the Fano peak as an interference effect involving resonant states. We find that there are three types of Fano asymmetry according to their origins: the interference between a resonant state and an anti-resonant state, that between a resonant state and a bound state, and that between two resonant states. We derive microscopic expressions of the Fano parameters that describe the three types of Fano asymmetry. We show that the last two types display the asymmetric energy dependence given by Fano, but the first one shows a slightly different form. 
%We also reveal that the Fano parameter of the first type can become complex under an external magnetic field.
%{\it Should we add a comment on the complex Fano paramater?}
\end{abstract}

\section{Introduction: Resonant States}
\label{HatanoSec1}

\subsection{Landauer formula and the transmission probability}
\label{HatanoSubsec1.1}

\begin{figure}[t]
\centering
\includegraphics*[width=0.65\textwidth]{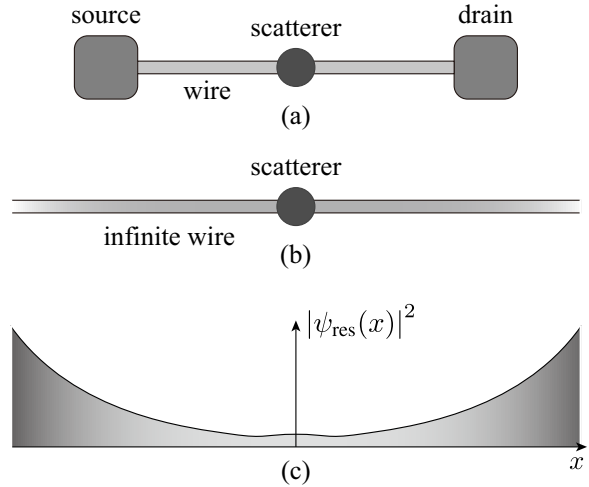}
\caption{(a) The situation of electronic conduction in mesoscopic systems. A quantum wire with a quantum scatterer connects a source and a drain.
The Landauer formula claims that the electronic conductance of the setup~(a) is proportional to the transmission probability of the setup~(b).
The spatial divergence~(c) of the resonant state of the scattering problem in the setup~(b) indicates that there are macroscopic numbers of electrons in the source and the drain in the setup~(a).}
\label{HatanoFig1}
\end{figure}

The Landauer formula~\cite{Landauer57,Datta95} tells us that the electronic conductance $\mathcal{G}$ in the situation of Fig.~\ref{HatanoFig1}~(a) (where we ignore the electron-electron interaction) is given by the transmission probability $T$ of the scattering problem in the infinite space as in Fig.~\ref{HatanoFig1}~(b):
\begin{align}
\mathcal{G}=\frac{2e^2}{h}T,
\end{align}
where $e$ is the elementary charge of an electron and $h$ is the Planck constant.
The two setups have the following common feature: once an electron goes out of the central quantum scatterer, it never comes back to the scattering area, at least not coherently~\cite{Hatano10}.
This is why the quantum scattering problem in the setup~(b) can describe the electronic conduction in the setup~(b).

The Fano asymmetry of the conductance is therefore equivalent to the Fano asymmetry of the transmission probability.
In the present chapter, we introduce the expansion of the transmission probability in terms of all eigenstates of the Hamiltonian with point spectra, including the bound, anti-bound, resonant, and anti-resonant states, but \text{not} including the scattering states with a continuous spectrum, thereby excluding the background integral~\cite{Sasada11,Klaiman11,Hatano10,Hatano11,Hatano13,Hatano14,Ordonez17a,Ordonez17b}.
Among the eigenstates with point spectra, the states other than the bound states, namely the resonant and anti-resonant states, mostly contribute to the electronic conduction; this is the point that we emphasize in the present chapter.
We describe the Fano asymmetry of the transmission probability in terms of interference between two point-spectral eigenvalues~\cite{Sasada11}.

We will try to make the chapter as self-contained as possible.
For the purpose, we will review in the present section the classification of the point-spectral eigenvalues in scattering theory in one dimension with a tutorial example.
In the next Section~\ref{HatanoSec2}, we will show the resonant-state expansion of the transmission probability with another tutorial example of the tight-binding model.
In Section~\ref{HatanoSec3}, we finally show our numerical analysis of the Fano asymmetry in the tight-binding model.
We find three types of the Fano asymmetry, depending on what pair of point-spectral states interferes with each other.

\subsection{Siegert boundary condition: a tutorial example}

\begin{figure}[t]
\centering
\includegraphics*[width=0.5\textwidth]{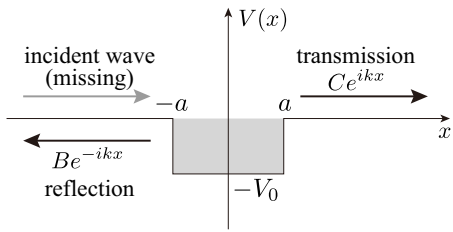}
\caption{The potential function given in Eq.~\eqref{HatanoEq30}. 
The discrete eigenstates are given by setting the incident wave to zero, which is the Siegert boundary condition.}
\label{HatanoFig2}
\end{figure}

We hereafter refer to the eigenstates with point spectra as the \textit{discrete} eigenstates.
The discrete eigenvalues are often defined as poles of the S-matrix, or the transmission amplitude in one dimension.
The transmission amplitude is given by the amplitude of the transmitted wave divided by that of the incident wave. 
Therefore, the poles of the transmission amplitude correspond to the zeros of the amplitude of the incident wave~\cite{Landau77,Hatano08,Hatano10}.
This means that the wave function of the discrete eigenvalues lacks the incident wave, retaining only the reflected and transmitted waves, in other words, only out-going waves; see Fig.~\ref{HatanoFig2}.
This is why the poles of the transmission amplitude are identified with the eigenstates of the time-independent Schr\"{o}dinger equation under the boundary conditions of out-going waves only, which was first set by Siegert~\cite{Siegert39}.
 
Let us demonstrate how to solve the time-independent Schr\"{o}dinger equation under the Siegert boundary condition.
We consider the standard equation
\begin{align}\label{HatanoEq20}
\left(-\frac{\hbar^2}{2m}\frac{d^2}{dx^2}+V(x)\right)\psi(x)=E\psi(x)
\end{align}
with the square-well potential
\begin{align}
V(x):=\begin{cases}
-V_0&\mbox{for $|x|<a$},\\
0&\mbox{for $|x|>a$},
\end{cases}
\end{align}
where $V_0>0$; see Fig.~\ref{HatanoFig2}.
We solve this under the Siegert boundary conditions
\begin{align}
\label{HatanoEq30}
\psi(x)\sim e^{ik|x|}.
\end{align}
More precisely, we assume the form
\begin{align}
\psi(x)=\begin{cases}
Be^{-ikx}&\mbox{for $x<-a$},\\
Fe^{ik'x}+Ge^{-ik'x}&\mbox{for $|x|<a$},\\
Ce^{ikx}&\mbox{for $x>a$},
\end{cases}
\end{align}
where
\begin{align}
\label{HatanoEq50}
E=\frac{\hbar^2k^2}{2m}=\frac{\hbar^2{k'}^2}{2m}-V_0.
\end{align}
We then set the connection conditions at $x=\pm a$, which produce four equations.
On the other hand, there are four unknown variables, namely the wave number $k$ (or equivalently the eigenenergy $E$) and the three ratios among the amplitudes $B$, $C$, $F$ and $G$.
We therefore obtain discrete solutions, namely point spectra.
Note that in finding the standard scattering states, we have another unknown variable, namely the amplitude of the incident wave $A$, in which case we obtain solutions for arbitrary $k$, namely continuous spectra.
This is the basic difference between the present discrete solutions and the continuum scattering solution.

\begin{figure}[t]
\centering
\includegraphics*[width=0.65\textwidth]{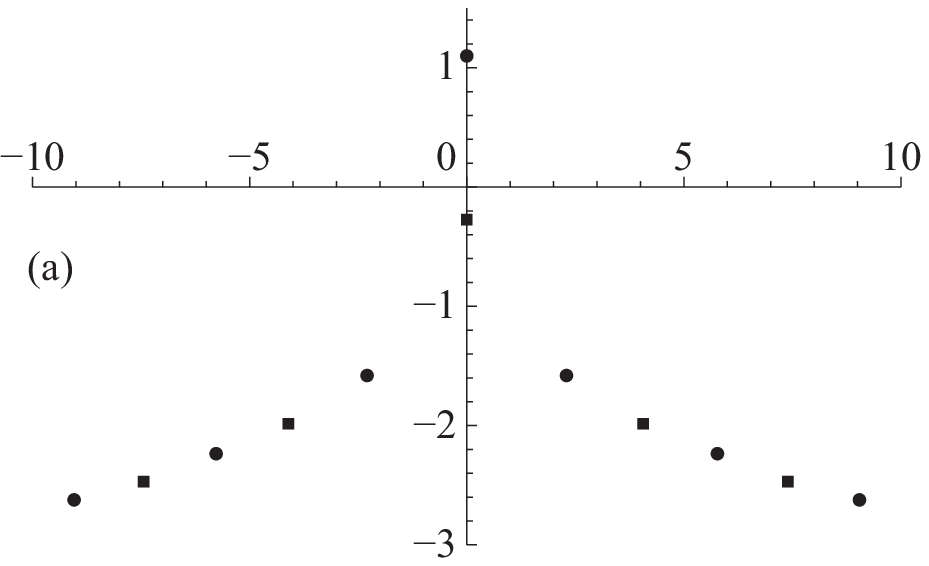}
\vspace{\baselineskip}
\\
\includegraphics*[width=0.65\textwidth]{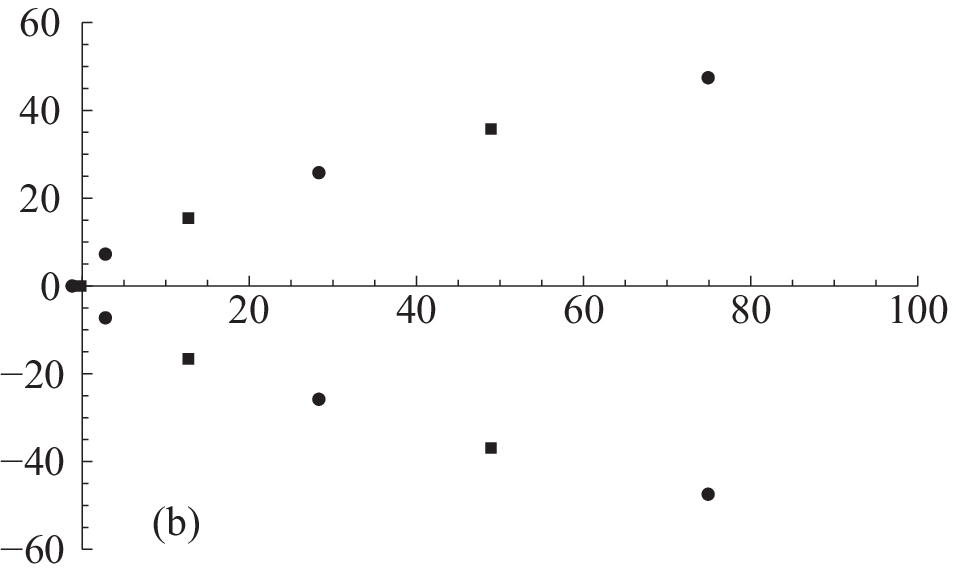}
\caption{The locations of the discrete eigenstates (a) in the complex-wave-number plane and (b) in the complex energy plane.
In both panels, the circles indicate the even solutions of Eq.~\eqref{HatanoEq90} and the squares the odd solutions of Eq.~\eqref{HatanoEq91}.}
\label{HatanoFig3}
\end{figure}

At this point, it is convenient to take advantage of the parity of the potential, finding even and odd solutions separately.
Even solutions should satisfy the equations $B=C$ and $F=G$, which yield
\begin{align}
\psi(x)=\begin{cases}
2F\cos(k'x)&\mbox{for $0<x<a$},\\
Ce^{ikx}&\mbox{for $x>a$}.
\end{cases}
\end{align}
The connection conditions give
\begin{align}
2F\cos(k'a)&=Ce^{ika},\\
2k'F\sin(k'a)&=ikCe^{ika}.
\end{align}
Dividing the second equation by the first one, we have
\begin{align}
\label{HatanoEq90}
k'\tan(k'a)=ik.
\end{align}
We obtain the even eigensolutions by solving Eqs.~\eqref{HatanoEq50} and~\eqref{HatanoEq90} simultaneously.
Similarly, odd solutions are given by solving Eq.~\eqref{HatanoEq50} together with
\begin{align}\label{HatanoEq91}
-k'\cot(k'a)=ik.
\end{align}
Numerically computing the solutions by the Newton-Raphson method, we obtain the solutions plotted in Fig.~\ref{HatanoFig3}.

\subsection{Resonant and anti-resonant states}

The state on the positive part of the imaginary axis of the complex wave-number plane is a bound state.
We can see this by inserting $k=i\kappa$ with $\kappa>0$ into the boundary condition~\eqref{HatanoEq30}.
It is located on the negative part of the real axis of the complex energy plane.

The states on the fourth quadrant of the complex wave-number plane are historically called resonant states,
while those on the third quadrant are called anti-resonant states.
These states are located, respectively, on the lower and upper halves of (the second Riemann sheet of) the complex energy plane.
We can prove that the time-reversal symmetry of the original problem~\eqref{HatanoEq20} dictates that each resonant state must have its complex conjugate partner of anti-resonant state;
although each of resonant and anti-resonant states breaks the time-reversal symmetry, the whole set of the solutions still observes the time-reversal symmetry.
The state on the negative part of the imaginary axis in the complex wave-number plane, and correspondingly on the negative part of the real axis in the second Riemann sheet of the complex energy plane, is called an anti-bound state, but we do not pay much attention to it throughout this chapter.

Since all the resonant and anti-resonant states are located on the lower half of the complex wave-number plane, their wave functions diverge spatially away from the scattering potential, being unnormalizable.
This is presumably the reason why they are often called unphysical.
Let us try in two ways to convince the readers that they are actually physical entities.

We first clarify a physical view of the spatial divergence~\cite{Hatano08,Hatano10} by multiplying the wave function~\eqref{HatanoEq30} by the temporal part as in
\begin{align}
\Psi_n(x,t)\sim e^{ik_n|x|-iE_nt}.
\end{align}
For the resonant states in the fourth quadrant, the real part of the eigen-wave-number $k_n$ is positive, while the imaginary part of the eigenenergy $E_n$ is negative. Therefore, the wave amplitude decays exponentially in time and the corresponding amount of the amplitude leaks towards positive and negative infinities.
For the anti-resonant states in the third quadrant, the real part of the eigen-wave-number is negative and the imaginary part of the eigenenergy is positive. Therefore, the wave amplitude comes into the central scattering area and the amplitude there grows exponentially in time.
The anti-resonant states are time-reversal of the resonant states.

\begin{figure}[t]
\centering
\includegraphics*[width=0.65\textwidth]{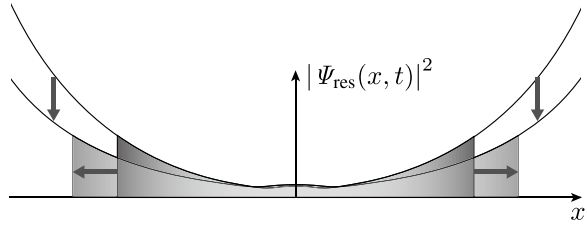}
\caption{The proof of the probability conservation.
As the time goes by, the wave function decays exponentially.
Accordingly, we expand the integration region to follow the exponential spatial divergence.}
\label{HatanoFig4}
\end{figure}

Based on this view, we can prove that the probability is conserved~\cite{Hatano08,Hatano10,Kawamoto11} under the following two conditions (Fig.~\ref{HatanoFig4}): first, we calculate the probability in a finite segment $[-L,L]$ containing the support $[-a,a]$ of the scattering potential; second, we let the integration area expand as in $[-L(t),L(t)]$ in order to chase the leaking amplitude for a resonant state. 
(We shrink the area for an anti-resonant state.)
The spatial divergence is exactly cancelled by the temporal decay.
This indicates that that spatial divergence is actually essential for the probability conservation.

Let us present another view of the spatial divergence.
In what we will show in Section~\ref{HatanoSec2}, we reveal that the transmission probability mostly comes from the resonant and anti-resonant states, in other words, the spatially divergent states.
This is indeed consistent with the original situation that was considered in the Landauer formula.
The Landauer formula assumes free electrons, neglecting electron-electron interactions, and hence we can regard the probability of the present one-electron problem as a quantity proportional to the number of electrons in a many-electron problem.
Therefore, the spatial divergence of the resonant and anti-resonant wave functions of the one-electron problem implies that there are macroscopic number of electrons far away from the scattering center.
As we see in Fig.~\ref{HatanoFig1}~(a), the system indeed has two baths (source and drain) away from the scattering potential, both of which have a macroscopic number of electrons under equilibrium.
These baths are the cause of the Joule heat generated by the resistance that the Landauer formula gives as the inverse conductance;
a microscopic number of electrons out of the source keeps the Fermi distribution of the source  during the energy-conserving quantum scattering all the way up until it meets a different Fermi distribution of the drain and is equilibrated to it. That is when the Joule heat is generated~\cite{Datta95}. 
In this view, it is essential for the Landauer formula to hold that the baths have a macroscopic number of electrons so that their Fermi distributions may never be disturbed by the microscopic number of electrons that participate in the conduction.
This situation is reproduced quantum-mechanically by the spatial divergence of the resonant and anti-resonant wave functions; see Fig.~\ref{HatanoFig1}~(c).

We thus stress that the spatially divergent resonant and anti-resonant states are not at all unphysical;
on the contrary, they are indispensable states for the electronic conduction.

\section{Resonant-state expansion: another tutorial example}
\label{HatanoSec2}

\subsection{Transmission probability and the Green's function}

\begin{figure}[t]
\centering
\includegraphics*[width=0.65\textwidth]{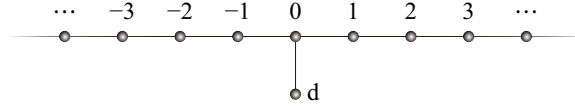}
\caption{The tutorial model of Eq.~\eqref{HatanoEq131}.}
\label{HatanoFig5}
\end{figure}

So far, we have considered a scattering potential in a continuum space.
We here move to a discretized model, namely the tight-binding model.
As another tutorial example, let us consider the T-shaped quantum-dot model (Fig.~\ref{HatanoFig5}):
\begin{align}\label{HatanoEq131}
H&:=-t_\textrm{hop}\sum_{x=-\infty}^\infty \left( |x+1\rangle \langle x|+|x\rangle\langle x+1|\right)
\nonumber\\
&+\varepsilon_0|0\rangle\langle0|
+\varepsilon_\mathrm{d}|\mathrm{d}\rangle\langle\mathrm{d}|
-g\left(|0\rangle\langle\mathrm{d}|+|\mathrm{d}\rangle\langle 0|\right),
\end{align}
where $t_\textrm{hop}$ is the hopping amplitude on the quantum wire with $-\infty<x<\infty$, $\varepsilon_0$ is the potential at the site $x=0$, $\varepsilon_\mathrm{d}$ is the potential at the quantum-dot site $\mathrm{d}$, and $g$ is the hopping amplitude between $x=0$ and $\mathrm{d}$.
We hereafter set $t_\textrm{hop}=1$ for brevity.
The central scattering area consists of the two sites $|\mathrm{d}\rangle$ and $|0\rangle$, while the rest is the environment.

As we emphasized in Subsection~\ref{HatanoSubsec1.1}, the electronic conductance is given by the transmission probability.
The energy dependence of the transmission probability of the model above is given by the Green's function in the following form~\cite{Fisher81}:
\begin{align}\label{HatanoEq141}
T(E)=(4-E^2)\left|\langle 0|\frac{1}{E-H}|0\rangle\right|^2.
\end{align}
The goal of the present section is to represent the Green's function in terms of all discrete states.
In order to do so, we first use the Feshbach formalism~\cite{Feshbach58} to eliminate the infinite number of the environmental degrees of freedom and express the Green's function by means of an effective Hamiltonian as small as a two-by-two matrix.
We then expand the Green's function of the effective Hamiltonian with respect to its eigenstates.

\subsection{Feshbach formalism for the tight-binding model}

Let us overview the Feshbach formalism~\cite{Feshbach58,Sasada11,Hatano14} here.
Our task is to find the solutions of the eigenvalue problem
\begin{align}
\label{HatanoEq130}
H|\psi\rangle=E|\psi\rangle.
\end{align}
It is difficult to solve it because $H$ is an $\infty$-by-$\infty$ matrix.
The Feshbach formalism gives an effective Hamiltonian for the central scattering area:
\begin{align}
\label{HatanoEq140}
H_\mathrm{eff}(E)\left(P|\psi\rangle\right)=E\left(P|\psi\rangle\right),
\end{align}
where 
\begin{align}
P:=|\mathrm{d}\rangle\langle\mathrm{d}|+|0\rangle\langle0|
\end{align}
with
\begin{align}
Q&:=I_\infty-P\nonumber\\
&=\sum_{x=-\infty}^{-1}
\left( |x\rangle \langle x-1|+|x-1\rangle\langle x|\right)
+\sum_{x=+1}^{+\infty}
\left( |x+1\rangle \langle x|+|x\rangle\langle x+1|\right).
\end{align}
Here $I_\infty$ denotes the identity operator in the entire space.
The effective Hamiltonian $H_\mathrm{eff}(E)$ in the present tutorial example is a two-by-two matrix.

Let us apply $P$ and $Q$ from the left of Eq.~\eqref{HatanoEq130} and at the same time insert $1=P+Q$ between $H$ and $|\psi\rangle$:
\begin{align}
PHP\left(P|\psi\rangle\right)+PHQ\left(Q|\psi\rangle\right)&=E\left(P|\psi\rangle\right),\\
QHP\left(P|\psi\rangle\right)+QHQ\left(Q|\psi\rangle\right)&=E\left(Q|\psi\rangle\right),
\end{align}
where we used the facts $P^2=P$ and $Q^2=Q$. We now eliminate $Q|\psi\rangle$ by finding
\begin{align}
Q|\psi\rangle=\frac{1}{E-QHQ}QHP\left(P|\psi\rangle\right)
\end{align}
from the second equation and inserting it into the first equation, obtaining
\begin{align}
\left(PHP+PHQ\frac{1}{E-QHQ}QHP\right)\left(P|\psi\rangle\right)&=E\left(P|\psi\rangle\right).
\end{align}
We thereby realize that the effective Hamiltonian in Eq.~\eqref{HatanoEq140} is given by
\begin{align}
\label{HatanoEq210}
H_\mathrm{eff}(E)=PHP+\Sigma(E)
\end{align}
with the `self-energy' 
\begin{align}
\label{HatanoEq220}
\Sigma(E):=PHQ\frac{1}{E-QHQ}QHP.
\end{align}

\subsection{Green's function of the effective Hamiltonian}

In order to calculate Eq.~\eqref{HatanoEq141}, we need the Green's function in the $P$ subspace:
\begin{align}
PGP=P\frac{1}{E-H}P.
\end{align}
We will prove here that this Green's function is equal to the Green's function of the effective Hamiltonian~\cite{Hatano14}:
\begin{align}
\label{HatanoEq241}
P\frac{1}{E-H}P=P\frac{1}{E-H_\mathrm{eff}(E)}P.
\end{align}

The proof is achieved by splitting the total Hamiltonian into the two parts 
\begin{align}
H=\left(PHP+QHQ\right)+\left(PHQ+QHP\right)
\end{align}
and carrying out the resolvent expansion:
\begin{align}
P\frac{1}{E-H}P&=P\frac{1}{E-\left(PHP+QHQ\right)}P\nonumber\\
&+P\frac{1}{E-\left(PHP+QHQ\right)}\left(PHQ+QHP\right)\frac{1}{E-\left(PHP+QHQ\right)}P\nonumber\\
&+P\frac{1}{E-\left(PHP+QHQ\right)}\left(PHQ+QHP\right)\frac{1}{E-\left(PHP+QHQ\right)}\nonumber\\
&\phantom{+}\times\left(PHQ+QHP\right)\frac{1}{E-\left(PHP+QHQ\right)}P
+\cdots
\end{align}
Using the fact $PQ=QP=0$, we find that only the even-order terms survive, obtaining
\begin{align}
P\frac{1}{E-H}P&=P\frac{1}{E-PHP}P\nonumber\\
&+P\frac{1}{E-PHP}PHQ\frac{1}{E-QHQ}QHP\frac{1}{E-PHP}P\nonumber\\
&+P\frac{1}{E-PHP}PHQ\frac{1}{E-QHQ}QHP\frac{1}{E-PHP}\nonumber\\
&\phantom{+}\times PHQ\frac{1}{E-QHQ}QHP\frac{1}{E-PHP}P
+\cdots\\
&=P\frac{1}{E-PHP}P+P\frac{1}{E-PHP}\Sigma\frac{1}{E-PHP}P\nonumber\\
&+P\frac{1}{E-PHP}\Sigma\frac{1}{E-PHP}\Sigma\frac{1}{E-PHP}P
+\cdots
\end{align}
Summing up the resolvent expansion with respect to $\Sigma$, we end up with~\cite{Hatano14}
\begin{align}
P\frac{1}{E-H}P&=P\frac{1}{E-\left(PHP+\Sigma\right)}P,
\end{align}
which is equivalent to Eq.~\eqref{HatanoEq241}.
This implies that we can expand the Green's function, and hence the transmission amplitude~\eqref{HatanoEq141}, with respect to the eigenstates of the effective Hamiltonian $H_\mathrm{eff}$.
This is what we will do in Sec.~\ref{HatanoSec2.4}.

\subsection{Calculation of the self-energy}

Before finding the eigenstates of the effective Hamiltonian, let us show an easy way~\cite{Sasada08} to compute the self-energy~\eqref{HatanoEq220}.
We can find it in a straightforward way~\cite{Sasada11,Hatano14} but there is a trick to compute it in a much easier way.
We first write down the Schr\"{o}dinger equation~\eqref{HatanoEq130} for $|x|\geq 1$:
\begin{align}
\label{HatanoEq230}
-\left(\psi_{x+1}+\psi_{x-1}\right)=E\psi_x,
\end{align}
where
\begin{align}
\psi_x:=\langle x| \psi\rangle.
\end{align}
Let us set the Siegert boundary conditions:
\begin{align}
\label{HatanoEq250}
\psi_x=Ce^{ik|x|}=C\times \begin{cases}
e^{-ikx}&\mbox{for $x\leq -1$},\\
e^{ikx}&\mbox{for $x\geq 1$}.
\end{cases}
\end{align}
Note that because of the discretized space, the real part of $k$ is limited to the first Brillouin zone $-\pi<\operatorname{Re}k<\pi$.
Inserting the wave function~\eqref{HatanoEq250} into Eq.~\eqref{HatanoEq230}, we have the dispersion relation
\begin{align}\label{HatanoEq351}
E=-2\cos k.
\end{align}
We also realize in the Schr\"{o}dinger equation for $|x|=1$ that $\psi_0=\langle 0 | \psi\rangle=C$.

We next write down the Schr\"{o}dinger equation for $x=0$ and for the dot site $\mathrm{d}$:
\begin{align}
-\left(\psi_{1}+\psi_{-1}\right)-g\psi_\mathrm{d}+\varepsilon_0\psi_0&=E\psi_0,\\
-g\psi_0+\varepsilon_\mathrm{d}\psi_\mathrm{d}&=E\psi_\mathrm{d}.
\end{align}
Inserting the Siegert boundary condition~\eqref{HatanoEq250} into the first equation, we have
\begin{align}
-2e^{ik}\psi_0-g\psi_\mathrm{d}+\varepsilon_0\psi_0=E\psi_0.
\end{align}
Combining this with the second equation, we can write down the matrix equation
\begin{align}
\label{HatanoEq300}
\begin{pmatrix}
\varepsilon_0-2e^{ik} & -g \\
-g & \varepsilon_\mathrm{d}
\end{pmatrix}
\begin{pmatrix}
\psi_0 \\
\psi_\mathrm{d}
\end{pmatrix}
=E
\begin{pmatrix}
\psi_0 \\
\psi_\mathrm{d}
\end{pmatrix}.
\end{align}
We identify~\cite{Sasada08} the two-by-two matrix on the left-hand side as the effective Hamiltonian $H_\mathrm{eff}$ in Eq.~\eqref{HatanoEq140}.
We can indeed confirm~\cite{Sasada11,Hatano14} Eq.~\eqref{HatanoEq210} with
\begin{align}\label{HatanoEq401}
PHP=
\begin{pmatrix}
\varepsilon_0 & -g \\
-g & \varepsilon_\mathrm{d}
\end{pmatrix},\qquad
\Sigma(E)=
\begin{pmatrix}
-2e^{ik} & 0 \\
0 & 0
\end{pmatrix}.
\end{align}
We can observe that the self-energy term functions as an effective complex potential at the site $x=0$, which makes the effective Hamiltonian non-Hermitian.

\subsection{Quadratic eigenvalue problem}
\label{HatanoSec2.4}

Let us stress here that the eigenvalue problem~\eqref{HatanoEq300} is not a standard one in the sense that the variable $k$, which is related to the energy $E$, exists on the left-hand side.
Therefore, this is a nonlinear eigenvalue problem.
In fact, we will show that this is formulated as a quadratic eigenvalue problem.

In order to solve the nonlinear eigenvalue problem, we introduce another energy-related variable as follows~\cite{Klaiman11,Hatano14}:
\begin{align}\label{HatanoEq411}
\lambda:=e^{ik}.
\end{align}
Because of the dispersion relation~\eqref{HatanoEq351}, the energy is given by
\begin{align}
E=-\left(\lambda+\frac{1}{\lambda}\right).
\end{align}
We can thereby transform Eq.~\eqref{HatanoEq300} into
\begin{align}
\begin{pmatrix}
\varepsilon_0-2\lambda & -g \\
-g & \varepsilon_\mathrm{d}
\end{pmatrix}
\begin{pmatrix}
\psi_0 \\
\psi_\mathrm{d}
\end{pmatrix}
=-\left(\lambda+\frac{1}{\lambda}\right)
\begin{pmatrix}
\psi_0 \\
\psi_\mathrm{d}
\end{pmatrix},
\end{align}
which is followed by
\begin{align}
\left[\lambda^2\begin{pmatrix}
-1 & 0 \\
0 & 1
\end{pmatrix}
+\lambda\begin{pmatrix}
\varepsilon_0 & -g \\
-g & \varepsilon_\mathrm{d}
\end{pmatrix}
+\begin{pmatrix}
1 & 0 \\
0 & 1
\end{pmatrix}
\right]
\begin{pmatrix}
\psi_0 \\
\psi_\mathrm{d}
\end{pmatrix}
=0.
\end{align}
The fact that the left-hand side is a second-order matrix polynomial of $\lambda$ is the reason why we call it a quadratic eigenvalue problem.
More formally, we have the following equation:
\begin{align}
\label{HatanoEq360}
Z(\lambda)\left(P|\psi\rangle\right)=0,
\end{align}
where
\begin{align}\label{HatanoEq361}
Z(\lambda):=\lambda^2(I_2 -\Theta)+\lambda PHP+I_2
\end{align}
with
\begin{align}
\Theta:=PHQHP
\end{align}
and $I_2$ is the two-dimensional identity operator.
We note 
\begin{align}
E-H_\mathrm{eff}(E)=-\frac{Z(\lambda)}{\lambda}.
\end{align}

There is a standard way of treating the quadratic eigenvalue problem~\cite{Tisseur01}.
We double the dimensionality as in~\cite{Klaiman11,Hatano14}
\begin{align}
\label{HatanoEq380}
\begin{pmatrix}
-\lambda  I_2 & I_2 \\
I_2 & \lambda(I_2-\Theta)+PHP
\end{pmatrix}
\begin{pmatrix}
P|\psi\rangle \\
\lambda P|\psi\rangle
\end{pmatrix}
=0.
\end{align}
Note that each matrix element is actually a two-by-two matrix.
The first row gives an equation that guarantees the doubled structure of the vector.
The second row gives the original quadratic eigenvalue problem~\eqref{HatanoEq360}.

\begin{figure}[t]
\centering
\includegraphics*[width=\textwidth]{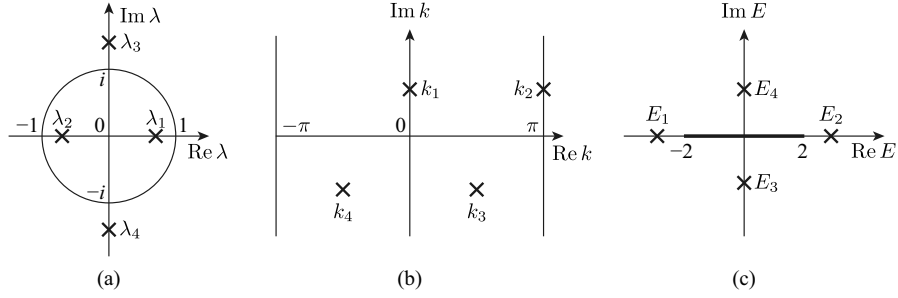}
\caption{A schematic view of the locations of the four discrete solutions (a) in the complex $\lambda$ plane, (b) in the complex-wave-number plane, and (c) in the complex energy plane.}
\label{HatanoFig6}
\end{figure}

We thereby find four solutions of the two-dimensional nonlinear eigenvalue problem~\eqref{HatanoEq300} out of 
\begin{align}
\det\begin{pmatrix}
-\lambda  I_2 & I_2 \\
I_2 & \lambda(I_2-\Theta)+PHP
\end{pmatrix}
=0,
\end{align}
which is a fourth-order equation with respect to $\lambda$.
In order to show the attributes of the solutions, let us solve the equation in the simplest case of $\varepsilon_0=\varepsilon_\textrm{d}=0$.
The equation is then reduced to
\begin{align}
\lambda^4+g^2\lambda^2-1=0,
\end{align} 
which produces
\begin{align}
\lambda^2=\frac{-g^2\pm\sqrt{g^4+4}}{2}.
\end{align}
The magnitude of the right-hand side is given by
\begin{align}
0<\frac{-g^2+\sqrt{g^4+4}}{2}<1,
\qquad
\frac{-g^2-\sqrt{g^4+4}}{2}<-1.
\end{align}
Therefore, 
\begin{align}\label{HatanoEq541}
\lambda_1,\lambda_2:=\pm\sqrt{\frac{-g^2+\sqrt{g^4+4}}{2}}
\end{align}
are located on the real axis inside the unit circle $|\lambda|=1$, while
\begin{align}\label{HatanoEq542}
\lambda_3,\lambda_4:=\pm i\sqrt{\frac{g^2+\sqrt{g^4+4}}{2}}
\end{align}
are located on the imaginary axis outside the unit circle $|\lambda|=1$; see Fig.~\ref{HatanoFig6}~(a).
This leads to the following solutions:
\begin{align}
k_1&:=\frac{i}{2}\left[\ln2-\ln\left(-g^2+\sqrt{g^4+4}\right)\right],
\\
k_2&:=\frac{i}{2}\left[\ln2-\ln\left(-g^2+\sqrt{g^4+4}\right)\right]+\pi,
\\
k_3&:=-\frac{i}{2}\left[\ln\left(g^2+\sqrt{g^4+4}\right)-\ln2\right]+\frac{\pi}{2},
\\
k_4&:=-\frac{i}{2}\left[\ln\left(g^2+\sqrt{g^4+4}\right)-\ln2\right]-\frac{\pi}{2}.
\end{align}
The former two solutions are located on the upper half of the complex wave-number plane, corresponding to bound states, while the latter two solutions are on the lower half, corresponding to a pair of resonant and anti-resonant states; see Fig.~\ref{HatanoFig6}~(b).
The only difference from the solutions in Fig.~\ref{HatanoFig3}~(a) of the continuous problem in Section~\ref{HatanoSec1} lies in the fact that a bound state $k=k_2$ exists on the line of $\operatorname{Re}k=\pi$ in addition to the one $k=k_1$ on the imaginary axis.

We finally obtain the eigenvalues of the problem~\eqref{HatanoEq130} in the form
\begin{align}
E_n=-\left(\lambda_n+\frac{1}{\lambda_n}\right).
\end{align}
For the two bound states, we have
\begin{align}
E_1,E_2=\mp\sqrt{2+\sqrt{g^4+4}},
\end{align}
which are located below and above the energy band $E=-2\cos k$.
For the resonant and anti-resonant states, we have
\begin{align}
E_3,E_4=\mp i\sqrt{-2+\sqrt{g^4+4}},
\end{align}
which are located on the imaginary axis of the second Riemann sheet of the complex energy plane; see Fig.~\ref{HatanoFig6}~(c).
Again, the only difference from the solutions in Fig.~\ref{HatanoFig3}~(b) is the existence of a bound state $E=E_2$ above the continuum (the energy band) in addition to the one $E=E_1$ below it.

In order to find the eigenvectors corresponding to the four eigenvalues, we cast Eq.~\eqref{HatanoEq380} into the form of the generalized linear eigenvalue problem:
\begin{align}
\label{HatanoEq460}
\left(A-\lambda B\right)|\Psi\rangle=0,
\end{align}
where
\begin{align}
A:=\begin{pmatrix}
0 & I_2 \\
I_2 & PHP
\end{pmatrix},
\qquad
B:=\begin{pmatrix}
I_2 & 0  \\
0 & \Theta-I_2
\end{pmatrix},
\qquad
|\Psi\rangle:=\begin{pmatrix}
P|\psi\rangle \\
\lambda P|\psi\rangle
\end{pmatrix}.
\end{align}
We can regard the generalized eigenvalue problem as the standard eigenvalue problem of the form
%\begin{align}
$\left(B^{-1}A-\lambda\right)|\Psi\rangle=0$
%\end{align}
as long as $B$ is invertible. 
Note that $B^{-1}A$ is an asymmetric matrix, although $A$ and $B$ are symmetric matrices.

%Let us briefly overview general properties of the generalized eigenvalue problem.
We denote the four right- and left-eigenvectors corresponding to the four eigenvalues $\lambda_n$ in Eqs.~\eqref{HatanoEq541}--\eqref{HatanoEq542} by $|\Psi_n\rangle$ and $\langle\tilde{\Psi}_n|$ with $n=1,2,3,4$.
The left-eigenvector $\langle\tilde{\Psi}_n|$ is not the Hermitian conjugate of the right-eigenvector $|\Psi_n\rangle$ because $B^{-1}A$ is an asymmetric matrix.
In order to fix the normalization of the eigenvectors, we take a look at the diagonal matrix element
\begin{align}
0=\langle \tilde{\Psi}_n|\left(A-\lambda_nB\right)|\Psi_n\rangle=\langle \tilde{\Psi}_n|A|\Psi_n\rangle
-\lambda_n\langle \tilde{\Psi}_n|B|\Psi_n\rangle.
\end{align}
We here normalize the eigenvectors so that $\langle\tilde{\Psi}_n|B|\Psi_n\rangle=1$,
which yields the eigenvalue in the form
\begin{align}
\langle \tilde{\Psi}_n|A|\Psi_n\rangle=\lambda_n.
\end{align}

We next check the orthogonality of the eigenvectors. 
For the purpose, we consider the matrix elements
\begin{align}
\langle \tilde{\Psi}_m|\left(A-\lambda_nB\right)|\Psi_n\rangle&=0,\\
\langle \tilde{\Psi}_m|\left(A-\lambda_mB\right)|\Psi_n\rangle&=0
\end{align}
for $m\neq n$.
Subtracting the second equation from the first one, we have
\begin{align}
\left(\lambda_m-\lambda_n\right)\langle\tilde{\Psi}_m|B|\Psi_n\rangle=0.
\end{align}
Assuming the lack of degeneracy, we have $\lambda_m\neq\lambda_n$, and hence $\langle\tilde{\Psi}_m|B|\Psi_n\rangle=0$, which is followed by $\langle\tilde{\Psi}_m|A|\Psi_n\rangle=0$.

To summarize, we have the diagonalization and the orthonormality
\begin{align}
\langle\tilde{\Psi}_m|A|\Psi_n\rangle&=\delta_{mn}\lambda_n,\\
\langle\tilde{\Psi}_m|B|\Psi_n\rangle&=\delta_{mn}
\end{align}
for general $m$ and $n$.
We can thereby expand the inverse of the four-by-four matrix $A-\lambda B$ in the form~\cite{Klaiman11,Hatano14}
\begin{align}
\frac{1}{A-\lambda B}=\sum_{n=1}^4|\Psi_n\rangle \frac{1}{\lambda_n-\lambda}\langle \tilde{\Psi}_n|.
\end{align}

\subsection{Resonant-state expansion of the Green's function}

The next task is to relate the inverse $(A-\lambda B)^{-1}$ to the Green's function~\eqref{HatanoEq241} so that we may expand the latter in terms of the discrete eigenstates $\lambda_n$.
We first block-diagonalize the matrix $A-\lambda B$ by means of the two matrices~\cite{Klaiman11,Hatano14}
\begin{align}
X(\lambda):=\begin{pmatrix}
-\lambda(I_2-\Theta)-PHP & I_2 \\
I_2 & 0
\end{pmatrix},
\qquad
Y(\lambda):=\begin{pmatrix}
I_2 & 0 \\
\lambda I_2 & I_2
\end{pmatrix}
\end{align}
as in
\begin{align}
X(\lambda)(A-\lambda B)Y(\lambda)=\begin{pmatrix}
Z(\lambda) & 0 \\
0 & I_2
\end{pmatrix},
\end{align}
where $Z(\lambda)$ was given in Eq.~\eqref{HatanoEq361}.
We therefore have the relation between the Green's function~\eqref{HatanoEq241} and the inverse $(A-\lambda B)^{-1}$ in the form~\cite{Klaiman11,Hatano14}
\begin{align}
P\frac{1}{E-H}P&=P\frac{1}{E-H_\mathrm{eff}(E)}P\nonumber\\
&=-\frac{\lambda}{Z(\lambda)}
=-\lambda \begin{pmatrix}
I_2 & 0 \\
\end{pmatrix}
\frac{1}{Y(\lambda)}\frac{1}{A-\lambda B}\frac{1}{X(\lambda)}
\begin{pmatrix}
I_2 \\ 0
\end{pmatrix}.
\end{align}
Since 
\begin{align}
\frac{1}{X(\lambda)}=\begin{pmatrix}
0 & I_2 \\
I_2 & \lambda(I_2-\Theta)+PHP
\end{pmatrix},
\qquad
\frac{1}{Y(\lambda)}=\begin{pmatrix}
I_2 & 0 \\
-\lambda I_2 & I_2
\end{pmatrix},
\end{align}
we arrive at the expansion of the Green's function~\eqref{HatanoEq241} as in~\cite{Klaiman11,Hatano14}
\begin{align}
P\frac{1}{E-H}P&=-\lambda 
\sum_{n=1}^4
\begin{pmatrix}
I_2 & 0 \\
\end{pmatrix}
|\Psi_n\rangle \frac{1}{\lambda_n-\lambda}\langle \tilde{\Psi}_n|
\begin{pmatrix}
0 \\I_2 
\end{pmatrix}
\nonumber\\
&=\sum_{n=1}^4
P|\psi_n\rangle \frac{\lambda\lambda_n}{\lambda-\lambda_n} \langle \tilde{\psi}_n|P.
\end{align}

Let us finally transform this to a more familiar form by using~\cite{Hatano14}
\begin{align}
E-E_n&=-\left(\lambda+\frac{1}{\lambda}-\lambda_n-\frac{1}{\lambda_n}\right)
%\nonumber\\
%&
=-(\lambda-\lambda_n)\left(1-\frac{1}{\lambda\lambda_n}\right)
\nonumber\\
&=\frac{\lambda-\lambda_n}{\lambda\lambda_n}\left(1-\lambda\lambda_n\right),
\end{align}
which is followed by~\cite{Hatano14}
\begin{align}
P\frac{1}{E-H}P&=\sum_{n=1}^4
P|\psi_n\rangle \frac{1-\lambda\lambda_n}{E-E_n} \langle \tilde{\psi}_n|P.
\end{align}
We can further show that the sum of the retarded and advanced Green's function
\begin{align}\label{HatanoEq680}
G^{R/A}:=\frac{1}{E-H\pm i\delta},
\end{align}
where $\delta$ is infinitesimally positive,
is given by~\cite{Sasada11}
\begin{align}\label{HatanoEq690}
P\Lambda P=\sum_{n=1}^4
P|\phi_n\rangle \frac{1}{E-E_n} \langle \tilde{\phi}_n|P,
\end{align}
where
\begin{align}\label{HatanoEq171}
\Lambda(E):=PG^R(E)P+PG^A(E)P=2\operatorname{Re}\left(PG^R(E)P\right)
\end{align}
for real $E$, while the states $|\phi_n\rangle$ and $\langle\tilde{\phi}|$ have different normalization from the states $|\psi_n\rangle$ and $\langle\tilde{\psi}|$ as in~\cite{Hatano14}
\begin{align}
|\phi_n\rangle:=\sqrt{1-{\lambda_n}^2}|\psi_n\rangle,
\qquad
\langle \tilde{\phi}_n|:=\sqrt{1-{\lambda_n}^2}\langle \tilde{\psi}_n|.
\end{align}

In Subsection~\ref{HatanoSubsec2.8}, we will express the transmission probability, and hence the electronic conductance, by means of $\Lambda$.
As we emphasized at the end of Sec.~\ref{HatanoSec1}, this expansion reveals that the electronic conduction is dominated by transmission through resonant and anti-resonant states;
the bound states in the expansion contribute to the transmission little.
Note again that the expansion~\eqref{HatanoEq690} does not contain scattering states.

%\subsection{Fisher-Lee formula of the transmission probability}
\subsection{New formula for the transmission probability}
\label{HatanoSubsec2.8}

We now come back to Eq.~\eqref{HatanoEq141} and derive a formula for the transmission probability in terms of the sum~\eqref{HatanoEq171}, which enables us to take advantage of the expansion~\eqref{HatanoEq690}.
We first show that the retarded and advanced Green's functions are given by the Green's function of the effective Hamiltonian $H_\mathrm{eff}(E)$ with $k$ in the self-energy term in Eq.~\eqref{HatanoEq401} set to be positive and negative, respectively.

We proved Eq.~\eqref{HatanoEq241} for general complex values of $E$.
We now set the energy $E$ to be real for the retarded and advanced Green's functions in Eq.~\eqref{HatanoEq680} to be used in the formula for the transmission probability.
We therefore consider the Green's function~\eqref{HatanoEq241} of the effective Hamiltonian $H_\mathrm{eff}(E)$ with real $E$.
Remember that the $E$ dependence of the effective Hamiltonian comes from the $k$ dependence of the self-energy in Eq.~\eqref{HatanoEq401}.
Since the energy $E$ and the wave number $k$ are related by the dispersion relation~\eqref{HatanoEq351}, two real values of $k$ give the same real value of $E$. 
The retarded Green's function corresponds to emission from a source at the origin, and hence has only out-going waves, while the advanced Green's function corresponds to absorption into a sink at the origin, and hence has only in-coming waves.
We therefore conclude~\cite{Sasada11,Hatano14} that the retarded Green's function is the Green's function of the effective Hamiltonian with positive $k$:
\begin{align}
PG^RP=P\frac{1}{E-H + i\delta}P=
\left.P\frac{1}{E-H_\mathrm{eff}(E)}P\right|_{0<k<\pi}.
\end{align}
Conversely, the advanced Green's function is that with negative $k$:
\begin{align}
PG^AP=P\frac{1}{E-H - i\delta}P=
\left.P\frac{1}{E-H_\mathrm{eff}(E)}P\right|_{-\pi<k<0}
\end{align}
with
\begin{align}
PG^A(E(k))P=PG^R(E(-k))P,
\qquad
PG^A(E(\lambda))P=PG^R(E(1/\lambda))P.
\end{align}

These give a formula for the Green's functions~\cite{Sasada11}:
\begin{align}\label{HatanoEq175}
i\Gamma(E)&:=
\left(PG^RP\right)^{-1}-\left(PG^AP\right)^{-1}
\\
&=-\left.\left[H_\mathrm{eff}(E(k))-H_\mathrm{eff}(E(-k))\right]\right|_{0<k<\pi}
\nonumber\\
&=\left.
\begin{pmatrix}
2(e^{ik}-e^{-ik}) & 0 \\
0 & 0
\end{pmatrix}
\right|_{0<k<\pi}
=\left.4i\sin k\right|_{0<k<\pi}|0\rangle\langle0|
\nonumber\\
&=2i\gamma|0\rangle\langle0|
\end{align}
for real $E$,
where
\begin{align}
\gamma:=\sqrt{4-E^2}.
\end{align}
More formally, it is written in the form
\begin{align}
\Gamma=\gamma PHQHP.
\end{align}
This function in conjunction with the retarded and advanced Green's functions is often used in the Fisher-Lee relation for the transmission probability~\cite{Fisher81}:
\begin{align}\label{HatanoEq770}
T(E)=\operatorname{Tr} \left(\Gamma G^R \Gamma G^A\right).
\end{align}
Since $\Gamma$ in our simple model has the only finite element for $|0\rangle\langle0|$, the Fisher-Lee relation~\eqref{HatanoEq770} is reduced to~\cite{Sasada11}
\begin{align}\label{HatanoEq790}
T(E)=\gamma^2\langle 0|G^R|0\rangle\langle 0|G^A|0\rangle
=\gamma^2\left|\langle 0|G^R|0\rangle\right|^2,
\end{align}
which is equivalent to Eq.~\eqref{HatanoEq141}.
We added a factor $1/4$ here because the two leads are attached to the same site in our simple model.

We now try to transform Eq.~\eqref{HatanoEq790} into an expression in terms of $\Lambda$ in Eq.~\eqref{HatanoEq171}.
For brevity, we denote $\langle 0 |G^R|0\rangle$, $\langle 0 |G^A|0\rangle$ and $\langle 0 |\Lambda|0\rangle$ by $G^R_{00}$, $G^A_{00}$ and $\Lambda_{00}$.
We solve 
\begin{align}\label{HatanoEq810}
\Lambda_{00}=G^R_{00} +G^A_{00}=2\operatorname{Re}G^R_{00},
\end{align}
which follows from Eq.~\eqref{HatanoEq171}, together with 
\begin{align}
G^A_{00}-G^R_{00}&=2i\gamma
G^R_{00} G^A_{00},
\end{align}
which follows from Eq.~\eqref{HatanoEq175}, or
\begin{align}\label{HatanoEq820}
\operatorname{Im}G^R_{00}=-\gamma\left|G^R_{00}\right|^2
=-\gamma\left[\left(\operatorname{Re}G^R_{00}\right)^2+\left(\operatorname{Im}G^R_{00}\right)^2\right].
\end{align}
Inserting Eq.~\eqref{HatanoEq810} into Eq.~\eqref{HatanoEq820}, we have
\begin{align}
\gamma \left(\operatorname{Im}G^R_{00}\right)^2
+ \operatorname{Im}G^R_{00}+\frac{ \gamma}{4}\left(\Lambda_{00}\right)^2 =0,
\end{align}
which produces
\begin{align}
-\gamma\left|G^R_{00}\right|^2
=\operatorname{Im}G^R_{00}&=
\frac{-1 \pm \sqrt{1- \left(\gamma\Lambda_{00}\right)^2}}{2\gamma}
\end{align}
The choice of the sign $\pm$ is given by the sign of
\begin{align}
\left|\varepsilon_0+\frac{g^2}{E-\varepsilon_\mathrm{d}}\right|-\gamma.
\end{align}
%depends on the relative sign of the eigenvalues of $\Lambda$, and hence cannot be fixed generally.
We finally arrive at~\cite{Sasada11}
\begin{align}\label{HatanoEq860}
T(E)=\frac{1}{2}\left[1\pm \sqrt{1-\left(\gamma\Lambda_{00}\right)^2}\right]
%=\frac{\gamma^2}{2}\frac{\left(\Lambda_{00}\right)^2}{1\mp \sqrt{1-\left(\gamma\Lambda_{00}\right)^2}}.
\end{align}

If the system is more complicated in such a way that two leads are attached to different sites of the central system, the formula becomes more complicated as in~\cite{Sasada11}
\begin{align}\label{HatanoEq920}
%T(E)&=\operatorname{Tr}\Xi\Xi^\ast
%=\operatorname{Tr}\Pi^2\left(\frac{1}{2}-i\alpha\right)\left(\frac{1}{2}+i\alpha\right)
%\nonumber\\
T(E)&=\left(\frac{1}{4}+\alpha^2\right)\operatorname{Tr}\Gamma \Lambda\Gamma \Lambda,
\end{align}
where
%The coefficients $\alpha$ and $\beta$ are given by representing everything in terms of the two eigenvalues of $\Pi$. After tedious but straightforward calculation, we arrive at
\begin{align}\label{HatanoEq1001}
\alpha^2=-\frac{1}{4}+\frac{1}{2(S^2-4D)}\left(4-D\pm\sqrt{(D+4)^2-4S^2}\right)
\end{align}
with
\begin{align}
S:=\operatorname{Tr}\Gamma\Lambda,
\qquad
D:=\operatorname{det}\Gamma\Lambda.
\end{align}
%This completes a proof of the new expression of the transmission probability, and hence the electronic conductance, in terms of $\Gamma$ and $\Lambda$ instead of $\Gamma$, $G^R$ and $G^A$.
See Ref.~\cite{Sasada11} for the derivation, including the choice of the sign in Eq.~\eqref{HatanoEq1001}.
%In terms of the new variable $\lambda$ defined in Eq.~\eqref{HatanoEq411}, the range $0<k<\pi$ for the retarded Green's function corresponds to the upper half of the unit circle in the complex $\lambda$ plane, while the range $-\pi<k<0$ for the advanced Green's function to the lower half.

\section{Fano asymmetry}
\label{HatanoSec3}

Since $\Lambda$ has the resonant-state expansion of the form~\eqref{HatanoEq690}, that is,
\begin{align}\label{HatanoEq1020}
\Lambda_{00}=\sum_{n=1}^4
\langle 0|\phi_n\rangle\frac{1}{E-E_n}\langle\tilde{\phi}_n|0\rangle,
\end{align}
we are ready to expand the transmission probability~\eqref{HatanoEq860} with respect to the discrete eigenstates.
For later use, let us split $\Lambda_{00}$ into the two parts: namely, the bound-state terms
\begin{align}\label{HatanoEq1030}
\Lambda^\textrm{b}(E)&:=\langle 0|\phi_1\rangle\frac{1}{E-E_1}\langle\tilde{\phi}_1|0\rangle
+\langle 0|\phi_2\rangle\frac{1}{E-E_2}\langle\tilde{\phi}_2|0\rangle,
\end{align}
and the resonant-anti-resonant-state terms
\begin{align}\label{HatanoEq1040}
\Lambda^\textrm{pair}(E)&:=\langle 0|\phi_3\rangle\frac{1}{E-E_3}\langle\tilde{\phi}_3|0\rangle
+\langle 0|\phi_4\rangle\frac{1}{E-E_4}\langle\tilde{\phi}_4|0\rangle.
\end{align}
We again emphasize that the latter produce the dominant contributions.

\begin{figure}[t]
\centering
\includegraphics*[width=0.75\textwidth]{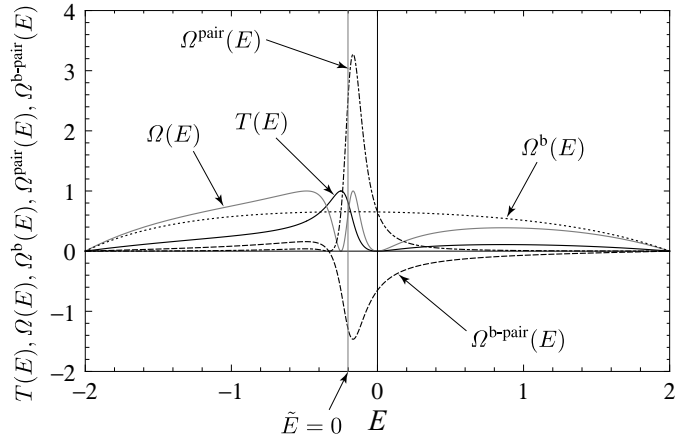}
\caption{The transmission probability $T(E)$ (black solid curve) as well as the quantities $\Omega(E)$ (gray solid curve), $\Omega^\textrm{b}(E)$ (dotted curve), $\Omega^\textrm{pair}(E)$ (chained curve), $\Omega^\textrm{b-pair}(E)$ (broken curve) for the parameter values $\varepsilon_0=4$, $\varepsilon_\mathrm{d}=0$ and $g=1$. The gray vertical line indicates the real part of the resonant and anti-resonant eigenvalues: $\operatorname{Re}E_3=\operatorname{Re}E_4=0.200606$.}
\label{HatanoFig7}
\end{figure}

Because we have the square of $\Lambda_{00}$ in the formula~\eqref{HatanoEq860}, there occur various interference terms.
We plot in Fig.~\ref{HatanoFig7} the following quantities:
\begin{align}
\Omega(E)&:=\left(\gamma\Lambda_{00}\right)^2
=\gamma^2\left(\Lambda^\textrm{b}(E)+\Lambda^\textrm{pair}(E)\right)^2,
\\
\Omega^\textrm{b}(E)&:=\left(\gamma \Lambda^\textrm{b}(E)\right)^2,
\\
\Omega^\textrm{pair}(E)&:=\left(\gamma \Lambda^\textrm{pair}(E)\right)^2,
\\\label{HatanoEq1141}
\Omega^\textrm{b-pair}(E)&:=\gamma^2 \Lambda^\textrm{b}(E)\Lambda^\textrm{pair}(E)
\end{align}
together with the transmission probability $T(E)$ as given in Eq.~\eqref{HatanoEq860}.
We can first observe that the transmission probability $T(E)$ (black solid curve) has a Fano asymmetric peak around the real part of the resonant and anti-resonant eigenvalues $E=\operatorname{Re}E_3=\operatorname{Re}E_4$ (gray vertical line), which results from the strong asymmetry of the quantity $\Omega(E)$ (gray solid curve).
We next observe that the bound-state contribution $\Omega^\textrm{b}(E)$ (dotted curve) is a smooth function, whereas the crossing term $\Omega^\textrm{pair}(E)$ (chained curve) between the resonant and anti-resonant states has a large peak around the energy of the pair and the crossing term $\Omega^\textrm{b-pair}(E)$ (broken curve) between the bound states and the pair of resonant and anti-resonant states also has a (negatively) large peak around the same point.
We can thereby conclude that the quantities $\Omega^\textrm{pair}(E)$ and $\Omega^\textrm{b-pair}(E)$ contribute to the asymmetry of $\Omega(E)$, and hence to the Fano asymmetry of $T(E)$.

Based on this observation, we classify the interference terms into three categories, which thereby result in three types of the Fano asymmetry~\cite{Sasada11}:
\begin{enumerate}
\renewcommand{\labelenumi}{(\roman{enumi})}
\setlength{\leftskip}{0.5em}
\item
Interference between a resonant state and the corresponding anti-resonant state;
\item
Interference between a bound state and a pair of resonant and anti-resonant states;
\item
Interference between two pairs of resonant and anti-resonant states.
\end{enumerate}
In the first case~(i), let us assume that for the resonant state $n=3$, the summand in Eq.~\eqref{HatanoEq1020} takes the form
\begin{align}
\langle 0|\phi_3\rangle\frac{1}{E-E_3}\langle\tilde{\phi}_3|0\rangle
=\frac{Ne^{i\theta}}{E-E_3}.
\end{align}
Note that the left-eigenvector $\langle\tilde{\phi}|$ is generally \textit{not} complex conjugate of the right-eigenvector $|\phi\rangle$, and hence the term above is generally complex.
Since the anti-resonant contribution $n=4$ is its complex conjugate, we have the term in Eq.~\eqref{HatanoEq1040} in the form
\begin{align}
\Lambda^\textrm{pair}(E)%&:=\langle 0|\phi_3\rangle\frac{1}{E-E_3}\langle\tilde{\phi}_3|0\rangle
%+\langle 0|\phi_4\rangle\frac{1}{E-E_4}\langle\tilde{\phi}_4|0\rangle
%\\
&=2\frac{N}{\left|\operatorname{Im}E_3\right|}\frac{\sin\theta+\tilde{E}\cos\theta}{1+\tilde{E}^2},
\end{align}
where 
\begin{align}
\tilde{E}:=\frac{E-\operatorname{Re}E_3}{\left|\operatorname{Im}E_3\right|}
\end{align}
is the normalized energy measure from the separation from $E=\operatorname{Re}E_3=\operatorname{Re}E_4$.
The sum of these two terms thereby contribute to the transmission probability $T(E)$ in the form~\cite{Sasada11}
\begin{align}\label{HatanoEq1060}
%T(E)
\Omega^\textrm{pair}
\sim\left(\frac{\tilde{E}+q^\textrm{pair}}{1+\tilde{E}^2}\right)^2,
\end{align}
where
\begin{align}\label{HatanoEq1142}
q^\textrm{pair}:=\tan\theta.
\end{align}
The peak of $\Omega^\textrm{pair}(E)$ around $\tilde{E}=0$ observed in Fig.~\ref{HatanoFig7} underscores the behavior in Eq.~\eqref{HatanoEq1060}.

The parameter~\eqref{HatanoEq1142} 
may be referred to as a Fano parameter, although Eq.~\eqref{HatanoEq1060} is not the original form derived by Fano~\cite{Fano61}:
\begin{align}\label{HatanoEq1080}
T(E)\sim\frac{\left(\tilde{E}+q\right)^2}{1+\tilde{E}^2}.
\end{align}
Indeed, many analyses take account only of resonant states, ignoring the corresponding anti-resonant states~\cite{Fano61,Sadreev03,Rotter09}.
This may be the reason why the behavior~\eqref{HatanoEq1060} has never been pointed out.
We will show below for the cases~(ii) and~(iii) that the Fano asymmetry in these cases take the form of Fano's formula~\eqref{HatanoEq1080}.
The reason why the denominator of Eq.~\eqref{HatanoEq1080} has a single power of $1+\tilde{E}^2$ in contrast to the double power in Eq.~\eqref{HatanoEq1060} is because the other states that interfere with the resonant state in question do not have singularities at $E=E_3$ nor at $E=E_4$.
Conversely, the double power of $1+\tilde{E}^2$ in the denominator of Eq.~\eqref{HatanoEq1060} is due to the fact that both the resonant and anti-resonant states have singularities with the same real part.
In other words, the new behavior~\eqref{HatanoEq1060} emerges only after we take account of the anti-resonant state in addition to the resonant state~\cite{Sasada11}.

Let us move to the second case~(ii).
This comes from the crossing term~\eqref{HatanoEq1141} in the square of $\Lambda_{00}$.
%also has a crossing term between the bound-state contributions $n=1,2$
%\begin{align}
%\langle 0|\phi_1\rangle\frac{1}{E-E_1}\langle\tilde{\phi}_1|0\rangle
%+\langle 0|\phi_2\rangle\frac{1}{E-E_2}\langle\tilde{\phi}_2|0\rangle
%\end{align}
%and the resonant--anti-resonant contributions $n=3,4$ in Eq.~\eqref{HatanoEq1040}.
We can derive an approximate energy dependence due to this interference by expanding it in terms of $\tilde{E}$, which
results in the form~\cite{Sasada11}
\begin{align}\label{HatanoEq1100}
%T(E)
\Omega^\textrm{b-pair}
\sim\frac{\left(\tilde{E}+q^\textrm{b-pair}\right)^2}{1+\tilde{E}^2}
\end{align}
for small $\tilde{E}$, where we can define the parameter $q^\text{b-pair}$ microscopically; see Ref.~\cite{Sasada11}.
We believe that this corresponds to Fano's phenomenological analysis~\cite{Fano61}.
The behavior~\eqref{HatanoEq1100} is indeed consistent with Fano's formula~\eqref{HatanoEq1080}.
The (negative) peak of $\Omega^\textrm{b-pair}$ around $\tilde{E}=0$ observed in Fig.~\ref{HatanoFig7} confirms Eq.~\eqref{HatanoEq1100}.

\begin{figure}[t]
\centering
\includegraphics*[width=0.65\textwidth]{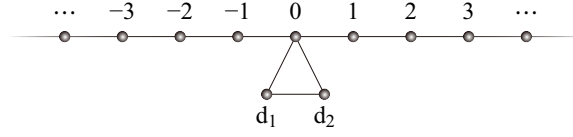}
\caption{The model~\eqref{HatanoEq1160}, which has two pairs of resonant and anti-resonant states in addition to two bound states.}
\label{HatanoFig8}
\end{figure}

In order to discuss the case~(iii), we need to move to a more complicated model that has multiple resonant states.
The simplest model with two pairs of resonant and anti-resonant states may be the following one~\cite{Sasada11}:
\begin{align}\label{HatanoEq1160}
H&:=-t_\textrm{hop}\sum_{x=-\infty}^\infty \left( |x+1\rangle \langle x|+|x\rangle\langle x+1|\right)
\nonumber\\
&+\varepsilon_0|0\rangle\langle0|
+\varepsilon_{\mathrm{d}_1}|\mathrm{d}_1\rangle\langle\mathrm{d}_1|
+\varepsilon_{\mathrm{d}_2}|\mathrm{d}_2\rangle\langle\mathrm{d}_2|
\nonumber\\
&
-g_{01}\left(|0\rangle\langle\mathrm{d}_1|+|\mathrm{d}_1\rangle\langle 0|\right)
-g_{02}\left(|0\rangle\langle\mathrm{d}_2|+|\mathrm{d}_2\rangle\langle 0|\right)
-g_{12}\left(|\mathrm{d}_2\rangle\langle\mathrm{d}_1|+|\mathrm{d}_1\rangle\langle \mathrm{d}_2|\right);
\end{align}
see Fig.~\ref{HatanoFig8}.
For an appropriate parameter set, it has two pairs of resonant and anti-resonant states in addition to two bound states below and above the energy band.
Let us denote the one pair of resonant and anti-resonant states by $n=3,4$ and the other pair by $n=5,6$.
The square of $\Lambda_{00}$ now has a crossing term between
\begin{align}
\Lambda^\textrm{pair}_1&:=\langle 0|\phi_3\rangle\frac{1}{E-E_3}\langle\tilde{\phi}_3|0\rangle
+\langle 0|\phi_4\rangle\frac{1}{E-E_4}\langle\tilde{\phi}_4|0\rangle
\end{align}
and
\begin{align}
\Lambda^\textrm{pair}_2&:=\langle 0|\phi_5\rangle\frac{1}{E-E_5}\langle\tilde{\phi}_5|0\rangle
+\langle 0|\phi_6\rangle\frac{1}{E-E_6}\langle\tilde{\phi}_6|0\rangle.
\end{align}

\begin{figure}[t]
\centering
\includegraphics*[width=0.75\textwidth]{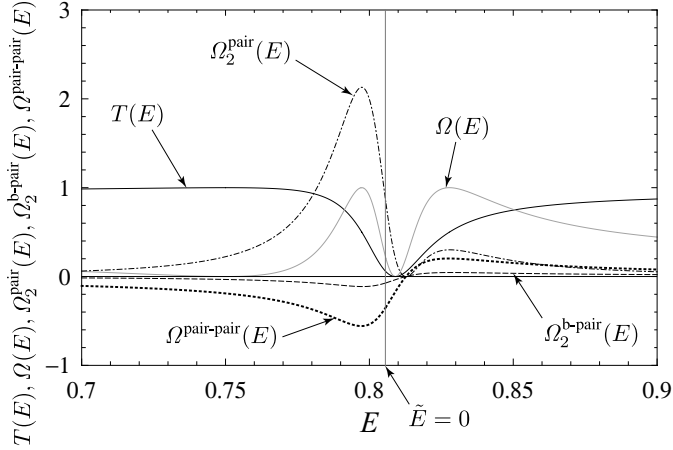}
\caption{The transmission probability $T(E)$ (black solid curve) as well as the quantities $\Omega(E)$ (gray solid curve), $\Omega^\textrm{pair}_2(E)$ (chained curve), $\Omega^\textrm{b-pair}_2(E)$ (broken curve), $\Omega^\textrm{pair-pair}(E)$ (thick dotted curve) for the parameter values $\varepsilon_0=\varepsilon_{\mathrm{d}_1}=0$, $\varepsilon_{\mathrm{d}_2}=1/2$ and $g_{01}=g_{02}=g_{12}=1/2$. The gray vertical line indicates the real part of the resonant and anti-resonant eigenvalues: $\operatorname{Re}E_5=\operatorname{Re}E_6=0.805784$.}
\label{HatanoFig9}
\end{figure}

We plot in Fig.~\ref{HatanoFig9} the following quantities:
\begin{align}
\Omega(E)&:=\left(\gamma\Lambda_{00}\right)^2
=\gamma^2\left(\Lambda^\textrm{b}(E)+\Lambda^\textrm{pair}_1(E)+\Lambda^\textrm{pair}_2(E)\right)^2,
\\
\Omega^\textrm{pair}_2(E)&:=\left(\gamma \Lambda^\textrm{pair}_2(E)\right)^2,
\\
\Omega^\textrm{b-pair}_2(E)&:=\gamma^2 \Lambda^\textrm{b}(E)\Lambda^\textrm{pair}_2(E)
\\
\Omega^\textrm{pair-pair}(E)&:=\gamma^2 \Lambda^\textrm{pair}_1(E)\Lambda^\textrm{pair}_2(E)
\end{align}
together with the transmission probability $T(E)$ as given in Eq.~\eqref{HatanoEq860}.
The other quantities not shown are smooth in the plotted region.
We observe that the interference between the two pairs of resonant and anti-resonant states, quantified by $\Omega^\textrm{pair-pair}$, is relatively large in this case.

This interference approximately results in the form~\cite{Sasada11}
\begin{align}
%T(E)
\Omega^\textrm{pair-pair}
\sim\frac{\left(\tilde{E}+q^\textrm{pair-pair}\right)^2}{1+\tilde{E}^2}
\end{align}
for small $\tilde{E}:=(E-\operatorname{Re}E_5)/|\operatorname{Im}E_5|$, where we can again define the parameter $q^\text{pair-pair}$ microscopically; see Ref.~\cite{Sasada11}.
Each of the two pairs affect the Fano asymmetry of the other pair, although the magnitudes of the Fano parameter can be very different from each other.
The interference between two resonant states has been discussed in Refs.~\cite{Sadreev03,Rotter09}, although we stress again that the corresponding anti-resonant states are mostly ignored.

The values of the Fano parameters around $E=0.805784$ are given by
\begin{align}
q^\textrm{pair}&=0.505055,
\\
q^\textrm{b-pair}&=-0.635981,
\\
q^\textrm{pair-pair}&=-0.607372
\end{align}
for $\varepsilon_0=\varepsilon_{\mathrm{d}_1}=0$, $\varepsilon_{\mathrm{d}_2}=1/2$ and $g_{01}=g_{02}=g_{12}=1/2$.
The signs of the three parameters indicate the parities of the Fano shapes.
The positive value of $q^\textrm{pair}$ is consistent with the fact that $\Omega^\textrm{pair}_2(E)$ has a peak on the left and a dip on the right, while the negative values of $q^\textrm{b-pair}$ and $q^\textrm{pair-pair}$ agree with the fact that both $\Omega^\textrm{b-pair}_2(E)$ and $\Omega^\textrm{pair-pair}(E)$ have a dip on the left and a peak on the right.

\section{Summary}
\label{HatanoSec4}

To summarize, we succeeded in expanding the transmission probability, and hence the electronic Landauer conductance, in terms of all discrete states but no continuous states~\cite{Sasada11,Klaiman11,Hatano14}.
This expansion makes more transparent to trace the cause of the Fano asymmetry back to the interference between various discrete states.
Fano's original argument~\cite{Fano61} considered the interference between a bound state and a resonant state, which produced the celebrated formula~\eqref{HatanoEq1080}.
We not only reproduced it but also found a new type of asymmetry with the double power in the denominator, which is caused by interference between a resonant state and its anti-resonant partner~\cite{Sasada11};
taking account of anti-resonant states made it possible.
We also reproduced the asymmetry due to the interference between two resonant-anti-resonant pairs.
We found microscopic derivation of the Fano parameters for the three types of the asymmetry~\cite{Sasada11}.
This may let us find experimentally the phase of a resonant state from $q^\textrm{pair}$ as in Eq.~\eqref{HatanoEq1142}, as well as from  $q^\textrm{b-pair}$ and $q^\textrm{pair-pair}$ .

We also found in Ref.~\cite{Sasada11} that the Fano parameter of the first type can become complex under an external magnetic field. 
This is consistent with experiments in Refs.~\cite{Kobayashi02,Kobayashi03,Kobayashi04}, which indeed observed complex Fano parameters.

\bibliographystyle{spphys}
\bibliography{hatano}
%
% Non-BibTeX users please follow the syntax
% the syntax of "referenc.tex" for your own citations
%\input{referenc}
%%%%%%%%%%%%%%%%%%%%%%%%%%%%%%%%%%%%%%%%%%%%%%%%%%%%%%%%%%%%%%%%%%%%%%  }

%%%%%%%%%%%%%%%%%%%%%%%%%%%%%%%%%%%%%%%%%%%%%%%%%%%%%%%%%%%%%%%%%%%%%%

\printindex
\end{document}